\documentclass{egpubl}

\JournalSubmission    
\usepackage[T1]{fontenc}
\usepackage{dfadobe}  

\usepackage{cite}  
\BibtexOrBiblatex
\electronicVersion
\PrintedOrElectronic
\ifpdf \usepackage[pdftex]{graphicx} \pdfcompresslevel=9
\else \usepackage[dvips]{graphicx} \fi

\usepackage{egweblnk} 
\usepackage{float}
\usepackage[table]{xcolor}
\usepackage{colortbl}
\usepackage{wrapfig}
\usepackage{soul}
\usepackage[mathletters]{ucs}
\usepackage[utf8x]{inputenc}
\usepackage{hyperref}
\usepackage{bm}
\usepackage{siunitx}
\usepackage[ruled,vlined,linesnumbered]{algorithm2e}
\usepackage{booktabs}
\usepackage{amsmath}
\usepackage{amssymb}
\usepackage{layouts}
\usepackage{lipsum}
\usepackage{mathptmx}
\usepackage{graphicx}

\definecolor{redcolor}{rgb}{0.8,0,0}
\definecolor{bluecolor}{rgb}{0,0,0.8}
\definecolor{greencolor}{rgb}{0,0.7,0}
\definecolor{browncolor}{rgb}{0.5,0.2,0.2}
\definecolor{greycolor}{rgb}{0.6,0.6,0.6}
\definecolor{purplecolor}{rgb}{0.6,0.0,0.6}

\usepackage[linesnumbered,ruled]{algorithm2e} 

\SetAlFnt{\small}
\SetAlCapFnt{\small}
\SetAlCapNameFnt{\small}
\SetAlCapHSkip{0pt}

\graphicspath{{./figs/}}

\newcommand{\refequ}[1]{Eq.~\ref{equ:#1}}

\DeclareOldFontCommand{\rm}{\normalfont\rmfamily}{\mathrm}
\def\argmin{\mathop{\rm argmin}}

\newcommand{\vc}[1]{\textbf{{$\mathsf #1$}}} 

\usepackage{cancel}
\newcommand{\q}{\ensuremath{{\vc q}}} 
\newcommand{\F}{\ensuremath{{\vc F}}} 

\newcommand{\bone}[1] {\ensuremath{\vc T}}

\newcommand{\Id} {\ensuremath{\vc I}}

\newcommand{\Q}{\vc{Q}}
\renewcommand{\d}{\vc{d}}
\renewcommand{\u}{\vc{u}}
\newcommand{\R}{\mathbb{R}}
\newcommand{\G}{\vc{G}}
\renewcommand{\S}{\mathbb{S}}
\newcommand{\EACAP}{E_\text{C}}
\newcommand{\Eobj}{E}
\newcommand{\dd}[2]{\frac{∂#1}{∂#2}}
\newcommand{\ddtwo}[2]{\frac{∂²#1}{∂#2²}}
\renewcommand{\H}{\vc{H}}
\newcommand{\B}{\vc{B}}

\newcommand{\g}{\vc{g}}

\title[EMU: Efficient Muscle Simulation in Deformation Space]%
      {EMU: Efficient Muscle Simulation in Deformation Space}

\author[V. Modi \& L. Fulton \& A. Jacobson \& S. Sueda \& D.I.W. Levin]
{\parbox{\textwidth}
	{\centering V. Modi$^{1}$\orcid{ 0000-0002-9350-494X}
     			, L. Fulton$^{1}$
     			, S. Sueda$^{2}$\orcid{0000-0003-4656-498X}
     			, A. Jacobson$^{1}$\orcid{0000-0003-4603-7143}
     			, D.I.W. Levin$^{1}$\orcid{0000-0001-7079-1934} 
        }
        \\
{\parbox{\textwidth}
{\centering $^1$University of Toronto, Toronto, Canada\\
         	$^2$ Texas A\&M University, College Station, TX}
}
}

\begin{document}

\teaser{
 \includegraphics[width=\textwidth, height=6cm]{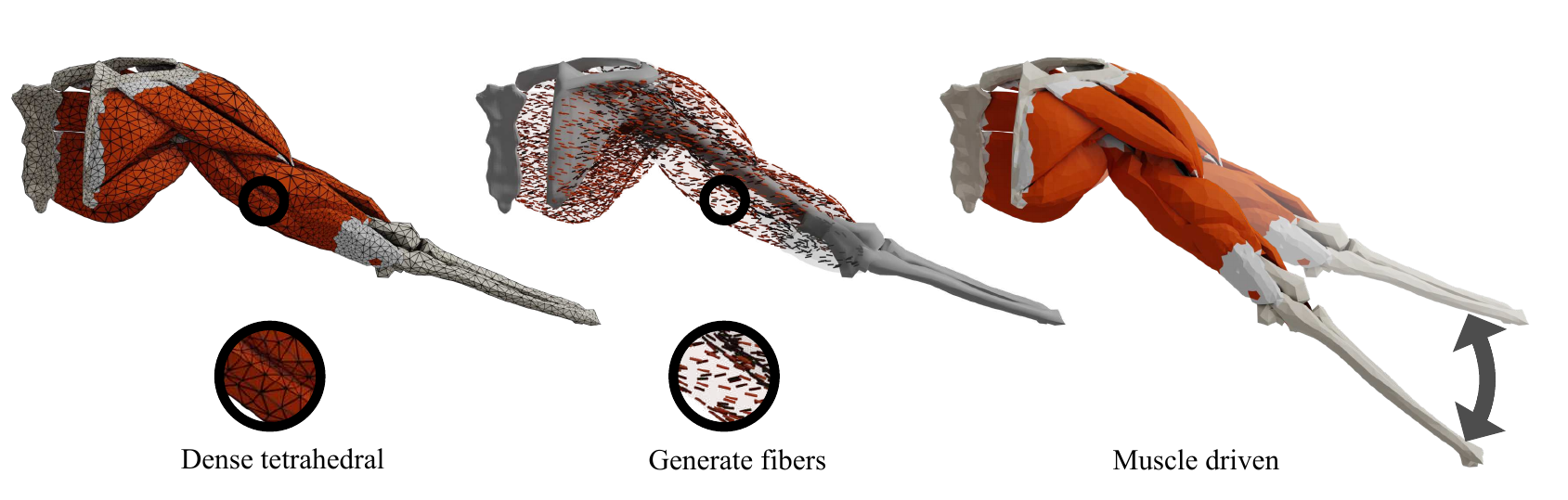}
    \centering
    \caption{ EMU allows volumetric muscle driven skeletal motion for efficient quasi-static simulation. EMU simulates volumetric musculoskeletal systems (left), complete with embedded, anisotropic fiber fields (middle), and correctly handles joints, stiff tendons, and bones to provide a holistic approach to musculoskeletal animation (right). EMU is asymptotically much faster than FEM---10x faster on a mesh size similar to the one above. (Video: 0m54s)}
    \label{fig:teaser}
}

\maketitle
\begin{abstract}
   EMU is an efficient and scalable model to simulate bulk musculoskeletal motion with heterogenous materials. First, EMU requires no model reductions, or geometric coarsening, thereby producing results visually accurate when compared to an FEM simulation. Second, EMU is efficient and scales much better than state-of-the-art FEM with the number of elements in the mesh, and is more easily parallelizable. Third, EMU can handle heterogeneously stiff meshes with an arbitrary constitutive model, thus allowing it to simulate soft muscles, stiff tendons and even stiffer bones all within one unified system. These three key characteristics of EMU enable us to efficiently orchestrate muscle activated skeletal movements. We demonstrate the efficacy of our approach via a number of examples with tendons, muscles, bones and joints. 

\ccsdesc[500]{Computing Methodologies ~ Applied computing}

\printccsdesc   
\end{abstract}

\section{Introduction}


An accurate portrayal of character motions in animation requires biologically representative musculoskeletal simulations. Computer graphics has a long and successful history of developing efficient simulations of elastica \cite{Terzopoulos1987}. However typical approaches suffer from both modeling and performance issues when applied to musculoskeletal applications. 
For instance, methods that rely on using a coarse simulation mesh also coarsen the muscle fiber field leading to difficulties generating realistic deformations, limitations in the ability to resolve small anatomical features such as tendons and numerical stiffening. Relying on fast, projective dynamics methods limits the types of material models that can be applied, which can lead to simulated behavior that is both visually off-putting and unstable. Finally, subspace methods coupled with optimized cubature can significantly alter the effect of material parameters making assets difficult to create as textbook material parameters cannot be used directly. In contrast, we provide an algorithm that can produces \textit{visually indistinguishable, unreduced} results that still scale well with the number of elements.

In this work we focus on quasi-static simulation of large-scale muscle activated motion. Quasi-static simulations, in which the inertial effects of the musculoskeletal system are ignored, are often used to animate movements. A well known example is Weta's Tissue solver which animates a large musculoskeletal motion as a series of small quasi-static steps. Quasi-statics is ideal to animate movements with slow-to-medium acceleration where secondary dynamics effects such as elastic wave propogation are visually unimportant. Time steps are essentially infinite, which leads to a variety of special problems such as tunneling effects during collision resolution.

Towards this goal, we propose an efficient finite element scheme which allows for the unified simulation of muscles, tendons, bones and joints, at high-resolution for both  bone- and muscle-first applications.  
We combine our simulator with a manual authoring system that allows a user to setup joints, build muscle fiber fields and identify tendon regions, given input surface geometry of a musculoskeletal system. Finally, we demonstrate the efficacy of our approach on a number of examples of musculoskeletal simulation.

To summarize, our method, EMU, offers the following three desirable attributes:
\begin{itemize}
  \item \textit{Visually accurate}: EMU produces results visually comparable to unreduced FEM.
  \item \textit{Efficient}: EMU scales and parallelizes well.
  \item \textit{Heterogeneous}: EMU simultaneously handles muscles, tendons, bones and joints in a unified fashion.
\end{itemize}

\section{Related Works}

A classical approach for bulk musculoskeletal (many muscles and bones together) simulation is to rely on standard finite element methods (FEM) applied to high resolution meshes in order to capture the required musculoskeletal detail as implemented in Weta's \textit{Tissue} software. However, such an approach is computationally intensive, suffers from numerical stiffening on heteregenous material, and is difficult to parallelize. Much of the subsequent research has focused on accelerating this procedure. Most approaches are stymied by three complicating factors:
\begin{enumerate}
\item The muscle constitutive model is complex, and so simple alternatives do not always provide robust, visually compelling results \cite{Smith2018}.
\item The motion of a muscle is heavily influenced by the embedded muscle fiber field. Using coarse meshes often leads to coarsening the fiber field as well, and this can therefore limit the space of actuated poses the muscle can reach \cite{Ichim2017}.
\item  The musculoskeletal system is heterogenous and composed of materials which exhibit wildly varying mechanical properties (tendon is $1000\times$ stiffer than muscle). These high stiffness ratios can cause numerical stiffening which can ``lock'' the system catastrophically \cite{Chen2017}.
\end{enumerate}
Below, we review previous attempts at tackling the important, but difficult problem of bulk musculoskeletal simulation. 

One approach is to represent the musculoskeletal system using line-of-force models. Here each muscle is represented not as a volume, but as a line (3D curve, potentially with wrapping surfaces or via points) along which a contractile force can act, and the skeletal system is represented as a system of articulated rigid bodies \cite{Lee2006,Delp2007,Sueda2008,Wang2012,Geijtenbeek2013,Lee2014}. Lines of force do not produce volumetric deformations or capture the richness of muscle fiber configurations and require coupling with rigid skeletons \cite{Sachdeva2015,Teran2003,Teran2005,Teran2005b,Lee2009,Si2014}.

Coarsening the simulation mesh can also yield speed-ups but at the cost of accurate deformations, as noted in Phace \cite{Ichim2017}. However muscle fiber fields must be ignored, or reformed via experimentation \cite{Levin2011}. Recently, fast projective approaches have been applied to muscle simulation \cite{Lee2018}. These approaches necessitate coarsening the simulation mesh and also restrict the class of constitutive models that can be applied. This can result in very stiff animations with artifacts \cite{Smith2018}. 

Eulerian methods have been used for musculoskeletal simulation~\cite{Fan2014} but these also eschew fiber field modeling and rely on kinematic coupling to rigidly simulated bones. Finally, data-driven approaches have also become popular \cite{Pons-Moll2015,Kim2017}, but these methods are designed for modeling the body as a whole and do not model muscles.

Reduction approaches such as substructuring and pose space methods are difficult to apply to bulk musculoskeletal simulation due to various locking problems that result from complex biomechanical setups \cite{Barbic2011,Xu2016}. Furthermore, reduction based methods limit material model choise and can significantly alter the behavior of nonlinear simulations \cite{BEH18,An2008,Fulton2019}. Sparse meshless methods~\cite{Faure2011} use frames and material-aware interpolation functions to significantly reduce the number of degrees-of-freedom in the numerical system. However, these methods have yet to demonstrate efficacy for large-scale, muscle-first simulation and often require approximating force evaluation to achieve good performance~\cite{Gilles2011}.

Multigrid methods promise exact solutions with linear scaling \cite{Brandt1977}, and have been applied to a number of computer graphics problems \cite{Zhu2010,Tamstorf2015}. However, heterogeneous nonlinear material, complex geometry, and time-varying activation present the worst-case scenario for constructing effective multigrid hierarchies. We avoid this and operate on a single volumetric mesh directly.

In this paper we focus on developing an unreduced and efficient algorithm for muscle-first simulation of musculoskeletal systems. The key to EMU's success is its use of deformation gradients, rather than nodal positions as the degrees-of-freedom of the simulation. This bares a resemblance to discontinuous methods for shape modeling \cite{Botsch2006,Kaufmann2008}. However, stiching the continuous mesh back together from discontinuous elements is still an open problem. An alternative discontinuous approach is rotation-strain coordinates which can unfortunately cause results to significantly differ from the gold-standard finite element approach~\cite{Pan2015}. 

EMU differs from previous discontinuous approaches by measuring discontinuity using the explicit minimizer of a coupling energy, rather than minimizing the coupling energy and physics energy of the system simultaneously. In this way, it has something in common with projective dynamics~\cite{Bouaziz2014,Ichim2017} or Fast Automatic Skinning Transforms~\cite{Jacobson2012}. However details matter, and there are key differences between EMU and projective dynamics as highlighted in \autoref{table:capabilities}. Although one component of our energy term resembles the term in Projective Dynamics, rather than minimize this energy using alternating projections or a variant of the alternating direction method of multipliers ~\cite{Narain2016}, we  leverage the algebraic properties of this energy to construct an efficient quasi-newton algorithm ~\cite{Wright1999} with capabilities \textit{beyond} the quasi-newton algorithms proposed in ~\cite{Liu2017, bcqn18, Li2019}.

\begin{figure}[h]
    \includegraphics[width=\columnwidth]{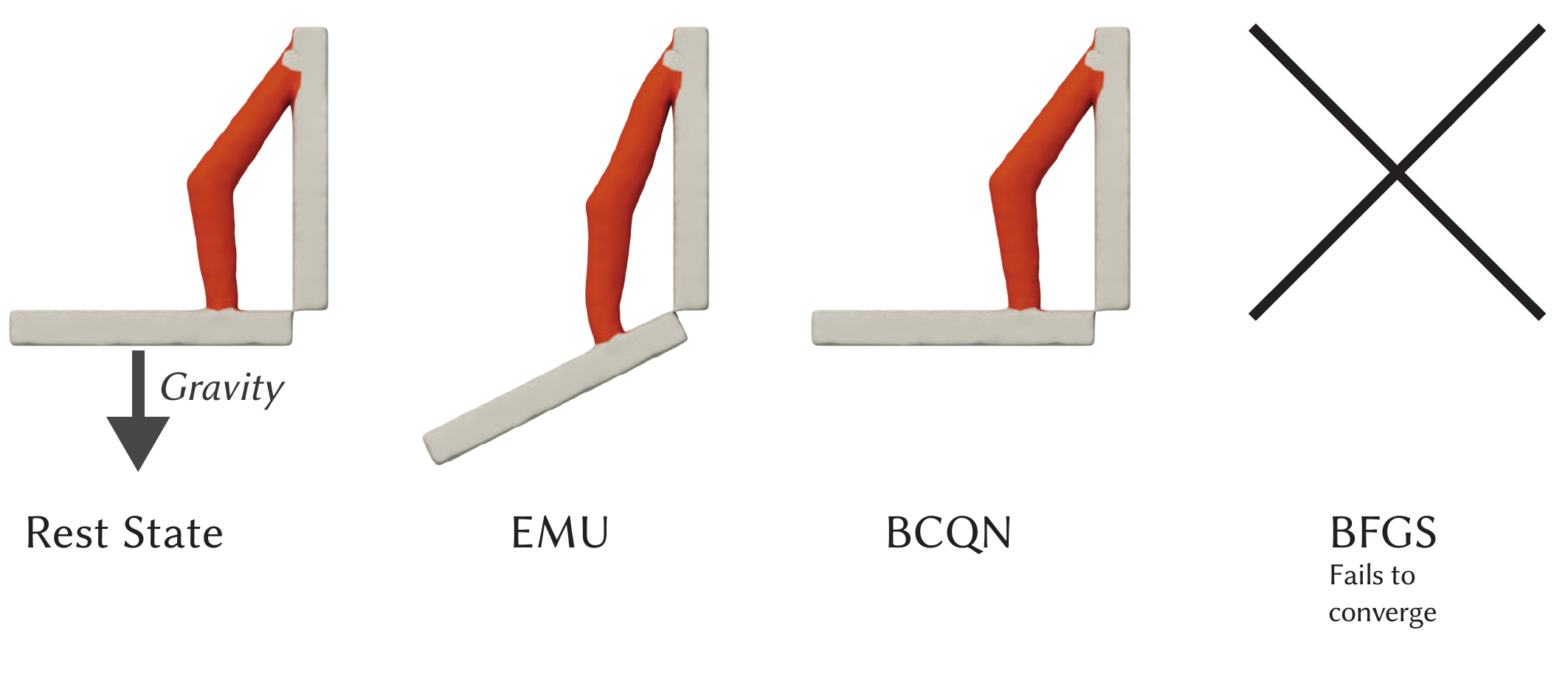}
    \caption{(Left) A simple, geometric hinge joint fixed at the top, under gravity. (Middle) With our framework, the hinge operates as expected. (Right) With BCQN the hingle locks.}
    \label{fig:emu_vs_bcqn}
\end{figure}

Readers should note that even though we describe EMU as a "quasi-newton method", we approximate our Hessian differently than other quasi-newton methods as described by \ref{section:quasi_newton_section}. Unlike EMU which uses an approximate Hessian to propagate second order information across joints, recent quasi-newton methods such as \cite{bcqn18} and \cite{Liu2017} use a BFGS like approach with a pre-conditioned Hessian which results in locking on the simplest meshes as shown in \autoref{fig:nmvsbfgs} and \autoref{fig:emu_vs_bcqn}. Our comparison in \autoref{fig:emu_vs_bcqn} shows the superiority of EMU for jointed systems. Other quasi-newton methods are applicable solely to homogenous, elastic, jello-like objects. Unlike EMU, they do not inherently support heterogenous materials and joints. Thus for \autoref{fig:emu_vs_bcqn}, the hinge joint is modeled geometrically as a shared edge between the two bone regions.

\begin{table*}[ht]
  \resizebox{\textwidth}{!}{
      \begin{tabular}{lccccccc} 
        \hline 
        \textbf{} 
        &  \multicolumn{1}{p{2cm}}{\centering Muscle Driven Motion }
        &  \multicolumn{1}{p{2cm}}{\centering Handles \\Joints }
        &  \multicolumn{1}{p{2cm}}{\centering Anisotropic Fibers }
        &  \multicolumn{1}{p{2cm}}{\centering Nonlinear Stiffness }
        &  \multicolumn{1}{p{2cm}}{\centering Handles Tendons }
        &  \multicolumn{1}{p{2cm}}{\centering No Coupling Terms }\\ [0.5ex] 
        \hline
        \rowcolor{black!20} 
        EMU                                  & Y & Y & Y & Y & Y & Y\\
        \cite{bcqn18} BCQN                   & N & N & N & Y & N & N\\ 
        \rowcolor{black!20} 
        \cite{Li2019} DOT                    & N & N & N & Y & N & N\\ 
        \cite{Liu2017} Liu et al.            & N & N & N & Y & N & N\\
        \rowcolor{black!20} 
        \cite{kim2019anisotropic} Kim et al. & N & N & Y & Y & N & N\\
        \cite{angles2019viper} VIPER         & N & Y & Y & N & N & N\\ 
        \rowcolor{black!20} 
        \cite{Lee2018} Lee et al.            & Y & N & Y & N & N & N\\
        [1ex]
        \hline
      \end{tabular}
  }
    \caption{Comparing the capabilities of EMU against some of its contemporaries. 'Y' indicates the capability is supported. 'N' indicates it is either not supported or not demonstrated in the paper. EMU alone supports all the features highlighted above. The column "No Coupling Terms" refers to the necessity of having extra coupling terms for rigid bones and soft muscles in other methods. EMU is the only method that handles muscles and bones in one unified framework.}
    \label{table:capabilities}
\end{table*}

\autoref{table:capabilities} summarizes the advantages of EMU over recent projection based methods. Of these methods, \cite{angles2019viper} and \cite{Lee2018} are the two most recent ones focused on musculoskeletal deformation. EMU's mesh density and muscle fiber density is much higher with noticeably better volume preservation. EMU handles tendons, which are three orders of magnitude stiffer than muscles. We handle joints and bones without any coupling terms. And we allow arbitrary constitutive models. EMU is a high performance algorithm with capabilities beyond those demonstrated by previous works.

\section{Method}
We model the bulk motion of a musculoskeletal system as a quasistatic elasticity
simulation driven by varying activation of muscle groups.
In this paper, we do not consider dynamics-dominated motions (e.g., leaping,
running, punching); instead we ignore inertial effects and focus on the
muscle-driven deformation of the musculoskeletal system by assuming slowly accelerating (but
non-trivial) activations.
In the language of continuum mechanics, this quasistatic deformation at some
time $t$ can be written as the solution to a scalar energy minimization
problem 
\begin{align}
 \begin{split}
  \label{equ:continuum}
  \argmin_{\q} ∫_{Ω} 
  Ψ_\text{iso}(  \F(\q)) +
  Ψ_\text{fiber}(\F(\q),\u,a(t)) -
  W(\q)
  \;d\Q,\\
  \text{subject to pin and joint constraints}
  \end{split}
\end{align}\\
over the domain $\omega$, where $\F : \omega → \R^{3x3}$ is the deformation gradient, 
$\q : \omega → \R^3$ is deformed positions of corresponding rest positions $\Q : Ω → \R³$ over the domain $Ω$. The internal potentials $Ψ_\text{iso}$ and
$Ψ_\text{fiber}$ are in general spatially varying, parameterized by materials, and defined as functions of the deformation gradient $\F$ which is based on $\q$.


In particular,
$Ψ_\text{iso}$ is a Neo-Hookean isotropic elastic potential constructed to be 
significantly stiffer in bone and tendon regions of the domain than in the muscles.
Meanwhile, $Ψ_\text{fiber}$ represents the active fiber-reinforcement
parameterized by the muscle fiber direction $\u : Ω → \S²$ and time-varying activation function
$a(t) : Ω×\R_{≥0} → \R$.
Work induced by external loads is captured by $W$, and we consider
constraints to the system such as pinning points, fixing regions (e.g., entire
bones) or constraining neighboring bones to rotate according to a specified
joint (see \autoref{sec:affine-bones}). Without loss of generality, we omit $W$ for the next section of this discussion and reintroduce it later on.

While we use stable Neo-Hookean materials ~\cite{Smith2018} and  a standard model of muscle behavior as a fiber reinforced
composite~\cite{Teran2003}, the discretization and numerical methods to follow accept any
valid potential energy for $Ψ_\text{iso}$ and $Ψ_\text{fiber}$.

Discretizing and optimizing the problem in \refequ{continuum} is numerically
daunting due to fiber anisotropy and the large disparity in material stiffness (bones, tendons, and muscles have Young's moduli of ${\sim}10^{10}$ Pa, ${\sim}10^{8}$ to ${\sim}10^{9}$ Pa and ${\sim}10^5$ Pa respectively as noted in ~\cite{maganaris1999vivo,rho1998mechanical}).
Discretizing bones and muscles differently requires awkward coupling constraints~\cite{Shinar2008} (e.g., treating bones as perfectly rigid objects and muscles as soft bodies). Even state-of-the-art coupling algorithms~\cite{Wang2019} require chain-rule-like computations to implement linearly-implicit time integration, which is significantly more complex than the non-linear statics solve that EMU performs.  
Meanwhile, direct simulation with the finite-element method suffers from numerical instability and poor convergence due to the high condition number of the resulting system. EMU's framework preserves the near perfect rigidity of real bones, stiffness of tendons and the compliance of soft muscles, without the overhead of any coupling constraints. 

\subsection{Discretization}
We propose discretizing the problem in \refequ{continuum} using a variable
separation approach.
Let $n$ and $m$ be the the number of vertices and tetrahedra, respectively.
We use vertex positions $\q ∈ \R^{3n}$ to track the volumetric deformation and
introduce \emph{independent variables} representing the deformation gradient
$\F_i ∈ \R^{3×3}$ for each tetrahedron, $i ∈ \{1,…,m\}$. 

For piecewise-linear finite-elements, the deformation gradient of a
tetrahedron is linearly \emph{dependent} on the deformed vertex positions:
\begin{align}
  \label{equ:defograd}
  \begin{pmatrix}
  \q_{i1}- \q_{i4} \\
  \q_{i2}- \q_{i4} \\
  \q_{i3}- \q_{i4}
  \end{pmatrix}^\top = 
  \F_i 
  \begin{pmatrix}
  Q_{i1}- Q_{i4} \\
  Q_{i2}- Q_{i4} \\
  Q_{i3}- Q_{i4}
  \end{pmatrix}^\top.
\end{align}
where $\q_{ij}$ and $Q_{ij}$ are the deformed and rest vertex positions of the
$j$th corner of the $i$th tetrahedron, respectively. 

However, we do \emph{not} require strict satisfaction of this equation.
Instead each independent deformation gradient $\F_i$ is free to
represent deformations of a much wider class of meshes than the continuous
tetrahedral mesh parametrized by vertex positions $\q$.
However, since we are ultimately interested in visualizing the continuous deformations, we find the \textbf{nearest continuous mesh}, $\q^*$, by satisfying \refequ{defograd} in a least-squares sense:
\begin{align}
\begin{split}
  \EACAP(\q,\F) & =
    \frac{1}{2} \sum\limits_{i=1}^m
    \left\| 
  \begin{pmatrix}
  \q_{i1}- \q_{i4} \\
  \q_{i2}- \q_{i4} \\
  \q_{i3}- \q_{i4}
  \end{pmatrix}^\top 
  \begin{pmatrix}
  Q_{i1}- Q_{i4} \\
  Q_{i2}- Q_{i4} \\
  Q_{i3}- Q_{i4}
  \end{pmatrix}^{-\top}
    -
  \F_i 
    \right\|² \\
  & =
    \frac{1}{2} \q^\top \G^\top \G \q - \q^\top \G^\top \F + \frac{1}{2} \F^\top\F\\
  \label{equ:ACAPEnergy}
\end{split}
\end{align}
where $\F ∈ \R^{9m}$ is a single vector stacking coefficients of all $m$ per-tet
deformation gradient \emph{variables} and $\G∈\R^{9m×3n}$ is the sparse matrix that
computes the \emph{actual} deformation gradients from the mesh deformed
according to $\q$.

This energy is zero when the deformed mesh implied by $\F$ is continuous. In other words, this energy describes the distance between our independent DOFs $\F$ and some continuous mesh represented by vertices $\q$. This amounts to a poisson-like solve with a constant Laplacian similar to ~\cite{yu2004mesh} where the poisson equation is viewed as an alternate to least-squares minimization. In our least-squares reconstruction of the nearest continuous mesh, the nullspace corresponding to rigid transformations is removed by fixing vertices in at least one bone. We refer to $\EACAP$ as the \emph{as-continuous-as-possible} (ACAP) energy.

Given a particular set of deformation gradients $\F$, the optimal deformed
vertex positions (or \textbf{nearest continuous mesh}) $\q^*$ that minimize $\EACAP$ are the solution to a sparse linear system:

\begin{equation}
  \label{equ:acap}
  \q^*= \argmin_\q \; \EACAP(\q,\F) = \left( \G^\top \G \right)^{-1} \G^\top \F.
\end{equation}

Out of all other continuous mesh representations $\q$, continuous mesh represented by $\q^*$ most closely resembles our deformation gradients $\F$. Any change in $\F$ would require us to re-solve the equation above for a new $\q^*$, thus making $\q^*$ a function of $\F$.

Since, $\q$ is now a function of $\F$, we can discretize the energy minimization \refequ{continuum} as a minimization over only $\F$:
\begin{equation}
  \min_\F \underbrace{Ψ_\text{iso}(\F) + Ψ_\text{fiber}(\F,\u,a(t)) + α \EACAP(\q(\F),\F)) - W(\q(\F))}_{\Eobj(\F)} \label{equ:separated}
\end{equation}
where $α≥0$ is a scalar parameter controlling the continuity implied by $\F$. Intuitively, we have replaced the hard constraint on continuity implied by standard finite element approaches, with a soft penalty approach, which will yield dividends later on.
For all finite choices of $α$, the deformation gradient variables can
\emph{break} continuity to provide a type of compliance in the system, which aids
optimization. A higher $\alpha$ generally means a higher level of continuity. As with all penalty methods, $\alpha$ can be chosen by the user to achieve a desired effect; however, in section \emph{3.7} we will detail an effective heuristic for choosing an $\alpha$ that provides both good simulation performance and visual accuracy.

With this variable separation we can see that the internal potential terms
$Ψ_\text{iso}$ and $Ψ_\text{fiber}$ in \refequ{separated} no longer depend on
the deformed vertex positions $\q$. Furthermore, if these are discretized using
the standard piecewise-constant strain assumption, these terms become \emph{easily parallelizable} summations over the tetrahedra. The computation at each tetrahedron only depends on data associated with exactly that tetrahedron (even the $\EACAP$ term is easily parallelizable as matrix-vector multiplications):
\begin{align}
  Ψ_\text{iso}(\F) & = \sum\limits_{i=1}^m Ψ_\text{iso}(\F_i), \\
  Ψ_\text{fiber}(\F,\u,a(t)) & = \sum\limits_{i=1}^m
  Ψ_\text{fiber}(\F_i,\u_i,a_i(t)).
\end{align}
In our examples, we use the (non-linear) inversion safe Stable Neo-Hookean energy \cite{Smith2018} for
$Ψ_\text{iso}$. Since biomechanical simulations do not require extreme elemental deformations, a non inversion-safe energy would work just as well as Stable Neo-Hookean as long as inversions are penalized by assigning a large (1e40) energy value to inverted elements.
For $Ψ_\text{fiber}$, we use linear activation \cite{Teran2003} along a
piecewise-constant direction field:
\begin{align}
  Ψ_\text{fiber}(\F_i,\u_i,a_i(t)) = a_i(t) \u_i^\top \F_i^\top \F_i \u_i,
\end{align} 
where $a_i(t) ≥0 $ is the non-negative activation at the $i$th tetrahedron at
time $t$, and $\u_i ∈ \S²$ is the unit-length fiber direction vector at the $i$th
tetrahedron.

\begin{figure}[H]
\includegraphics[width=\columnwidth]{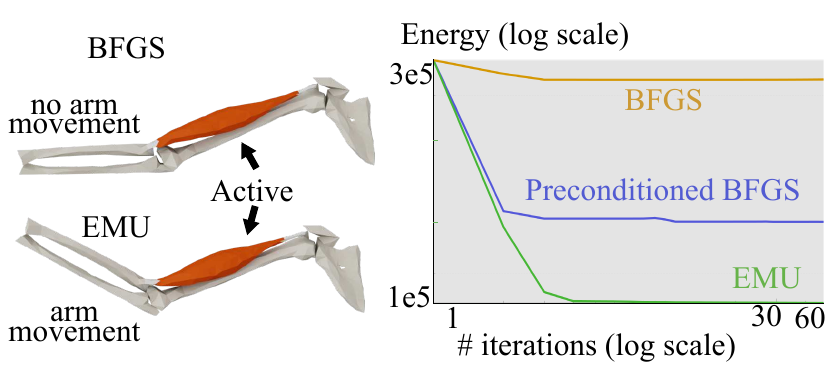}
\caption{ BFGS approximates the Hessian using only gradient information. This leads to catastrophic locking at joints. Our method works as expected.} 
\label{fig:nmvsbfgs}
\end{figure}

\subsection{Descent-direction solver choice}

Next we turn to the question of how to best minimize
\refequ{separated}. One option is BFGS. BFGS~\cite{Wright1999} or its limited-memory variant (LBFGS) are quasi-newton methods that are both popular and effective for physics simulation \cite{BATHE1980}. These approaches have the benefit of only requiring the gradient of the objective function, avoiding expensive Hessian computations. The gradient of \refequ{separated} can be computed as:
\begin{equation}
\begin{split}
  \frac{d\Eobj}{d\F} &=  
    \dd{Ψ_\text{iso}}{\F} + 
    \dd{Ψ_\text{fiber}}{\F} + 
    α \frac{d\EACAP}{d\F}\\
  &= \dd{Ψ_\text{iso}}{\F} + 
    \dd{Ψ_\text{fiber}}{\F} + 
    α \dd{\EACAP}{\F} + 
    α \cancelto{0}{\dd{\EACAP}{\q}}\dd{\q}{\F},
    \label{equ:gradient}
\end{split}
\end{equation}
where
\begin{align}
\label{equ:gradient2}
    \dd{\EACAP}{\F} &=
    \G \q(\F) + \F = 
    - \G \left( \G^\top \G \right)^{-1} \G^\top \F + \F.
\end{align}
Here we utilize the optimality of $\q(\F)$ to eliminate the term depending
on $∂\EACAP/∂\q$. This $d\Eobj/d\F$ can be computed efficiently by \emph{precomputing} a sparse 
factorization of the constant sparse symmetric matrix $\G^\top \G ∈ \R^{3n×3n}$.

Unfortunately, our experiments showed that the BFGS method can get immediately
stuck in a locked configuration. This is especially likely once we later introduce
joint constraints. Because the physics energies are completely decoupled ($Ψ_\text{iso}$ and $Ψ_\text{fiber}$ depend only on $\F_i$ of each tetrahedron), it becomes the job of
$\EACAP$ to distribute motion throughout the system. At the initial position,
both $\EACAP$ and $∂\EACAP/∂\F$ are $0$. Upon activation of a muscle, the individual muscle tetrahedra contract, but there is no movement induced in the bones. Because bones and tendons are very stiff relative to the muscles, the system will remain in place, unable to transfer the force created by muscle contraction across the joints (\autoref{fig:nmvsbfgs}). Essentially, this force transfer is a second-order effect and is not captured by the gradient, which only provides information about each tetrahedron in isolation. Previous work shows BFGS is not well suited for this type of application \cite{bcqn18}. Even pre-conditioning the LBFGS search direction does not sufficiently capture the second-order effects needed for joint motion.

To incorporate these second-order effects, the natural solution would be to simply use newton's method. However standard newton's method requires the computation of the Hessian matrix 
\begin{align}
  \label{equ:full-Hessian}
  \frac{d^2\Eobj}{d\F^2} &= 
  \ddtwo{Ψ_\text{iso}}{\F} + \ddtwo{Ψ_\text{fiber}}{\F} + α \ddtwo{\EACAP}{\F},\\
  \intertext{with}
  \ddtwo{\EACAP}{\F} &=
    - \G \left( \G^\top \G \right)^{-1} \G^\top + \Id.
\end{align}
Terms $\ddtwo{Ψ_\text{iso}}{\F}$ and $\ddtwo{Ψ_\text{fiber}}{\F} $
are sparse, block diagonal and simple to compute. However, the last term ($\ddtwo{\EACAP}{\F}$) requires taking the inverse of a sparse matrix (which is dense), making a direct computation of the Hessian intractable for all the but smallest of examples. For example, on a mesh with ~50k tets, the dense inverse would be 1GB large. Many of our examples are similar in size or larger. Thus, even construction of this dense Hessian is prohibitory, let alone computation with it. Therefore, standard newton's method is not viable for EMU.

\begin{figure}[H]
\includegraphics[width=\columnwidth]{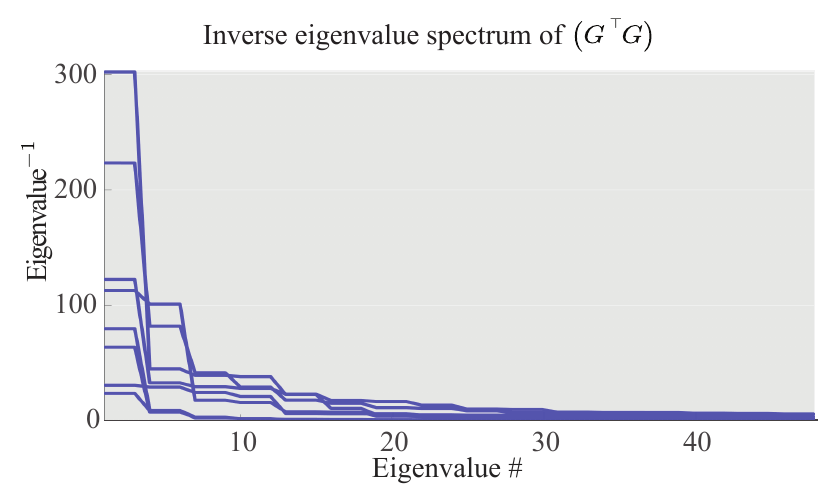}
\caption{Eigenvalue spectrum of $G^\top G$ for all the examples. Regardless of the example mesh, the first 48 eigen-modes capture most of the variance in $G^\top G$. Thefore we choose  a number of eigen vectors that encompasses 90 percent of the spectrum.}
\label{fig:eigenvaluespectrum}  
\end{figure}

\subsubsection{Alternative Quasi-Newton Method}
\label{section:quasi_newton_section}
Instead we derive a new quasi-newton approach, similar in asymptotic performance to the standard BFGS update. A quasi-newton method is one which doesn't use the exact Hessian ~\cite{Wright1999}. Motivated by this we derive a fast new way to evaluate approximate Hessian which uses the exact Hessian for the first two sparse, block-diagonal terms ($∂²Ψ_\text{iso}/∂\F²$ and $∂²Ψ_\text{iso}/∂\F²$) but approximates the last, dense term $∂²\EACAP/∂\F²$ using a low-rank approximation.

The dominant step in \refequ{full-Hessian} is the inversion of the large matrix $\G^\top \G$.
Therefore, we take the eigen decomposition:
\begin{align}
  \left( \G^\top \G \right)^{-1} \approx D = Φ^\top Λ^{-1} Φ,
  \label{equ:low-rank}
\end{align}
where $Φ∈\R^{k×3n}$ collects the eigen vectors corresponding to the lowest $k$ eigen values, which are placed along the diagonal of $Λ∈\R^{k×k}$.
Observing the eigenvalue spectrum of $G^\top G$ (\autoref{fig:eigenvaluespectrum}), we justify a reduction via the first $k<<m$ eigen-modes. In this case, we notice that $k=48$ sufficiently describes over $90\%$ of the variance in $G^\top G$ for all our examples.

Substituting \refequ{low-rank} in \refequ{full-Hessian} results in an expression
that still does not yet admit efficient \emph{solving} with the Hessian, which is now:
\begin{align}
  \frac{d²\Eobj}{d\F²} &\approx
    \H - α \B^\top Λ^{-1} \B,
  \label{equ:not-yet}
\end{align}
where
\begin{align}
  \H &= \ddtwo{Ψ_\text{iso}}{\F} + \ddtwo{Ψ_\text{fiber}}{\F} + α\Id
    \quad\text{ and }\quad
  \B = Φ\G^\top.
\end{align}
The matrix $\H ∈ \R^{9m×9m}$ is composed of $9×9$ blocks along the diagonal.
A key insight now becomes apparent. We can make use of the Woodbury matrix
identity (see, e.g., \cite{James1999AAR}) which holds that:
\begin{align}
\left(A + UCV \right)^{-1} = 
  A^{-1} - A^{-1}U \left(C^{-1} + VA^{-1}U \right)^{-1} VA^{-1},
\end{align}
for correctly sized matrices $A,U,C,V$ (and $A,C$ invertible).

Applying the Woodbury matrix identity to the inverse of the Hessian expression
in \refequ{not-yet} produces:
\begin{align}
  (\frac{d²\Eobj}{d\F²})^{-1} & \approx 
  \H^{-1} + \alpha\H^{-1}\B^\top\left(Λ - \alpha\B\H^{-1}\B^\top \right)^{-1} \B\H^{-1}.
  \label{equ:Hessian-woodbury}
\end{align}
Each iteration of our quasi-newton solver will need to multiply this expression
on the right with the gradient $d\Eobj/d\F$ from \refequ{gradient} to determine
the step direction.
We can compute this \emph{action} very efficiently:
to solve against $\H$ we precompute a $9×9$ dense factorization
corresponding to each tetrahedron and conduct back substitutions in parallel.
We also use this action to compute the dense 
$k×k$ matrix 
$Λ + \alpha\B\H^{-1}\B^\top$ in \refequ{Hessian-woodbury} and then solve against it (e.g., using a factorization
method at run-time). Multiplications against the precomputed $k×9m$ dense
rectangular
matrices $\B$ are conducted in parallel using Eigen \cite{eigenweb}.

Our Hessian approximation is guaranteed to be symmetric-positive-definite (SPD) since each term is SPD (we use the standard definiteness fix for the elastic energy). Therefore, our quasi-newton search direction (derived via a low rank approximation of \textbf{only} the dense term) is guaranteed to be a \textit{descent} direction.

Once the step direction is computed, we use a back-tracking line search, satisfying the Armijo condition, to ensure sufficient decrease in the objective so our method converges to a local minimum. 
Unlike modal reduction methods (e.g., \cite{Xu2016}) which permanently alter the
solution space, we only use  the eigendecomposition to build an approximate Hessian and retain the exact gradient. Importantly, our quasi-newton optimization approximates the search direction, but solves the \emph{full-space} problem in \refequ{separated}. 

\begin{figure}[H]
\includegraphics[width=\columnwidth]{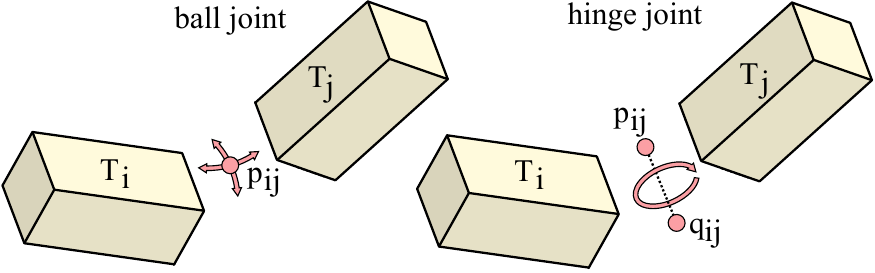}
\caption{Joints are implemented as algebraic constraints on the vertices of corresponding bones. We employ two types of joints. Ball joints (left) and hinge joints (right). }
\label{fig:joints}  
\end{figure}

\subsection{Affine Bones}\label{sec:affine-bones}
Armed with an efficient solver for muscles, we turn our attention to adding bones to the system. Remarkably, our method allows us to model bones, muscles and tendons in one system without any coupling terms. Due to their stiffness, bones deform negligibly and so, we chose to represent all the $\F_i$ for a bone mesh by a single deformation gradient. We model bone rigidity by applying a very high young's modulus to the bone elements and ensure the rigidity of the deformation gradient translates to rigid motion of the bone vertices through a constraint as shown in ~\refequ{acap-constraints} by the constraint $B_q q = B_F F$. In this equation $B_fF$ are the deformation gradients of the bones elements and $B_q q$ represents the deformation gradients of the bones on the continuous mesh.

\begin{align}
    \q^*= \argmin_\q \; \EACAP(\q,\F) = \left( \G^\top \G \right)^{-1} \G^\top \F, \\
    \text{s.t.} \qquad B_q q = B_F F\\
         Jq = 0
  \label{equ:acap-constraints}
\end{align}

In \refequ{acap}, rather than represent the position of bones using their vertex positions, we instead use a single affine transform for each bone. Joint constraints can then be represented as affine constraints $Jq = 0$ added to \refequ{acap-constraints} and shown geometrically in \autoref{fig:joints}. The joint constraint expresses that for each joint connecting two bones, the deformation of bone one and bone two, as applied to the points of the joint, are the same, thus $Jq = 0$. This yields a KKT system, shown in \refequ{kkt-matrix}, where the left-hand-side is a constant. Though, theoretically, $\epsilon_1 = \epsilon_2 = 0$, in practice however, both the joint constraint and bone constraints apply to the same DOFs. Therefore, we introduce a little slack into one of them. By setting $\epsilon_1 = -1e-4$ we prevent numerical problems while ensuring the deformation remains virtually unaffected.
\begin{align}
  \begin{pmatrix}
    \G^T \G & J^T & B^T \\
    J & \epsilon_1 & 0 \\
    B & 0 & \epsilon_2 \\
  \end{pmatrix}
  \begin{pmatrix}
   q \\
   \lambda_1 \\
   \lambda_2 \\
  \end{pmatrix}
  = 
  \begin{pmatrix}
   \G^T F \\
   0 \\
   B_F F \\
  \end{pmatrix}
  \label{equ:kkt-matrix}
\end{align}

For best performance we use the parallelizable Pardiso solver. Iterative methods are not competetive since the left-hand-side can be pre-factored (we tested Conjugate Gradients and found it to be slower). Thus EMU is able to handle muscle, tendons, bones and joints all within one contiguous system without the need for coupling terms.  In practice we find that it is useful to incorporate weights proportional to material stiffness into \refequ{acap} for muscles and tendons, but not for bones. For bones,  this is akin to adding a very stiff spring to the physical system which can lead to locking.

\begin{algorithm}[ht]
  \SetAlgoLined
  \SetKwRepeat{Do}{do}{while}
  \KwData{\F - Initial guess for deformation gradients}
  \KwResult{\F - Updated deformation gradients}
  //Line search step size

  $\sigma \gets 10 $

  //Line search f tolerance

  $c \gets 10^{-4}$

  //Line search decrement

  $\rho \gets 0.5$

  \Do{($\|\mathbf{g} \| < \epsilon_1 \; or \; (\Psi(F) - e_i) < \epsilon_2 $)}{

    //Calculate initial energy

    $e_i = \Psi(\F)$

    //Calculate the gradients

    $ \mathbf{g} \gets \frac{d\Psi(\F)}{dF} $
    
    //Woodbury method to find the search direction

    $ \mathbf{d} \gets \H(\F)^{-1}\mathbf{g} $

    //Backtracking line search to find step size

    \Do{($\Psi(\F_{temp}) < e_i + \sigma c \g^\top \d $)}{

      //Update temp F with descent direction

      $\mathbf{F_{temp}} = \F + \sigma \mathbf{d}$

      //ACAP solve from \refequ{acap-constraints}

      $\q = argmin_{\q} \; E_c(\q, \mathbf{F_{temp}})$

      $\sigma \gets \rho\cdot\sigma$
    }

    //Update F with new descent direction

    $\F \gets \F + \sigma*\mathbf{\mathbf{d}}$
    
    //Vertex-wise gravity force on the continuous vertices. 
    $Mg$ is the full mass matrix times the per-vertex gravity accelerations.
    
    $\mathbf{v} \gets Mg$ 

    //At first set external work gradient (negative external force) to gravity by \refequ{external-forces}

    $\mathbf{g_{ext}}$ = ExternalForces$(\mathbf{v})$ 

    //Compute contact forces that counteract mesh overlap.

    \Do{($\|\delta \mathbf{f_c}\| < \epsilon_{contact}$)}{

        //Run ~\cite{heidelberger2004consistent} on the continuous vertices

        $\mathbf{f_c}$ = CollisionForces$(\q)$

        //Update external work gradient

        $\mathbf{g_{ext}} \gets \mathbf{g_{ext}} - \mathbf{f_c}$
        
        //Woodbury method to update the search direction with contact forces

        $\mathbf{d_{contact}} \gets \H(F)^{-1}\mathbf{g_{ext}} $

        //Update temp F with descent direction
        
        $\mathbf{F_{temp}} = \F + \mathbf{d_{contact}} $

        //ACAP solve from \refequ{acap-constraints}
        
        $\q = argmin \; E_c(\q, \mathbf{F_{temp}})$
    }

    //Update F with descent direction accounting for external force 
    
    $\F \gets \F + \mathbf{d_{contact}}$
    
  }

 \label{algo:collide}
 \caption{Our contact-aware iterative newton solver.}
\end{algorithm}

\begin{figure}[H]
    \includegraphics[width=\linewidth]{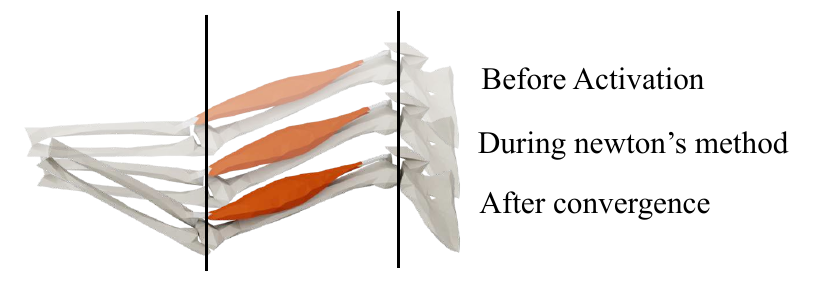}
    \caption{ During newton's method, we update the constraints on \refequ{acap} in order to introduce compliance between the bone $F$ and the bone's vertices. Notice the length of the bone temporarily changes within the quasi-newton method allowing the muscle elements to attain a plausible deformation without locking the system. After convergence, \refequ{acap} is run once more with updated constraints that ensure the bone $F$ exactly match the continuous mesh, thus verifying that the bone does not deform.}
    \label{fig:muscle-ruler}
\end{figure}

Additionally, we find that in order to prevent locking while increasing the quality of the deformation, certain liberties must be taken with joint and bone deformation constraints in \refequ{acap-constraints}. Bones on the continuous mesh must be allowed to slightly deviate from the discontinuous elements during newton's method in order for the algorithm to find a good search direction as shown in \autoref{fig:muscle-ruler}. This requires a loosening of the bone deformation constraint in \refequ{kkt-matrix} during newton's method by setting $\epsilon_2 = -1e-3$ while tightening the joint constraint by setting $\epsilon_1 = 0$. After the method has converged, the constraint can simply be updated $\epsilon_2 = 0$ to ensure strict adherence of the bone vertices to the bone's deformation gradient and introducing a negligible slack on the joint as $\epsilon_1 = -1e-4$.

\subsection{External Forces}

External forces such as gravity can be applied to our system using the standard Jacobian transpose method for converting per-vertex forces acting on the continuous tetrahedral mesh to generalized forces acting on the per-tetrahedron deformation gradients. The rate of work done by an external force is given by 

\begin{equation}
  \dot{\q}^T\bm{f}_{ext} = \dot{\F}^\top\underbrace{\left(\left( \G^\top \G \right)^{-1} \G^\top\right)^\top\bm{f}_{ext}}_{\mbox{Generalized Force: } \nu}.
  \label{equ:external-forces}
\end{equation}

For constant forces, such as gravity, the work in \autoref{equ:continuum} becomes $\F^\top\nu$ where $\nu$ can be efficiently computed at startup using the prefactored $\G^\top \G$. During the optimization, $f_{ext}$ is added as a constant external force.

\subsection{Collision Resolution}

Although not the focus of this work, the advantage of using a quasi-newton search strategy for optimization is that we can easily incorporate standard collision resolution into the algorithm --- all that is required is a method of: (1) detecting collisions in between muscles and bones and (2) computing forces that will resolve these collisions. For (1) and (2) we rely on \cite{heidelberger2004consistent}. We run collision resolution after the line search in our quasi-newton method to ensure that our meshes are collision free at the end of each newton iteration, similar to other projection based algorithms for contact and constraint handling. In \autoref{algo:collide}, we show the details of our particular implementation. We exploit the efficiency of inverting the EMU Hessian and our prefactored ACAP Hessian to propagate per-vertex contact forces through the mesh efficiently. These two properties can be exploited in other contact-aware gradient based algorithms as well. We typically allow a small amount of overlap in our simulations as it helps to reach converged solutions in cases with many closely conforming muscles. In general contact handling in such scenarios remains an open problem in graphics that we leave for future explorations.

It should be noted that although we show, for the benefit of the reader, that standard collision resolution methods work within EMU, this interpenetration is biologically infeasible. Muscle and bones are surrounded by sheaths of connective tissue called fascia, which limit deformation and limit contact. The simplest and most biologically accurate solution for interpenetrating muscles would be to fuse the muscle meshes together and activate each section separately.

\subsection{Modeling}
Our musculoskeletal models start life as separate triangle meshes for each bone and muscle. In all our examples, we must manually set joint locations for each socket and hinge joint. Next, we fuse muscle and bone meshes by manually overlapping them and then tetrahedralizing using TetWild~\cite{Hu2018,Sellan2018}. After this stage, each tetrahedron is labelled as either muscle or bone. We manually select muscle tetrahedra near the origin and insertion of each muscle to serve as tendons. Finally we assign material properties to our tetrahedral mesh. For muscles we use Young's Modulus of $6e6$ Pa, bones $1e10$ Pa and tendons $4.5e8$ Pa to $1.2e9$ Pa derived from biological measurements. We use a Poisson's ratio of $0.49$ for all materials. 
\begin{figure}[H]
  \includegraphics[width=\linewidth]{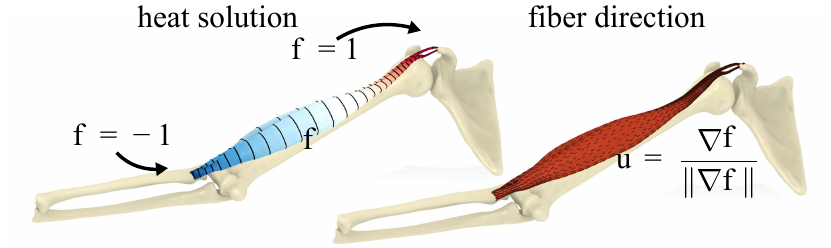}
  \caption{The muscle fiber directions $\u$ are computed as the
    gradient direction of a solution to the heat equation $∆f=0$ with sources
    and sinks at opposite attachment points.}
  \label{fig:muscle-direction-field} 
\end{figure}
For each muscle, we automatically compute a fiber direction $\u$ using the heat equation. We set Dirichilet boundary conditions of $1$ and $-1$ at insertion and origin points of the muscle and compute the equilibrium heat distribution. We take the normalized gradient of this field to be the fiber direction shown in \autoref{fig:muscle-direction-field}.

\subsection{The Weighting Paramter} \label{sec:alpha-paramter}
\begin{figure}[H]
    \includegraphics[width=\linewidth]{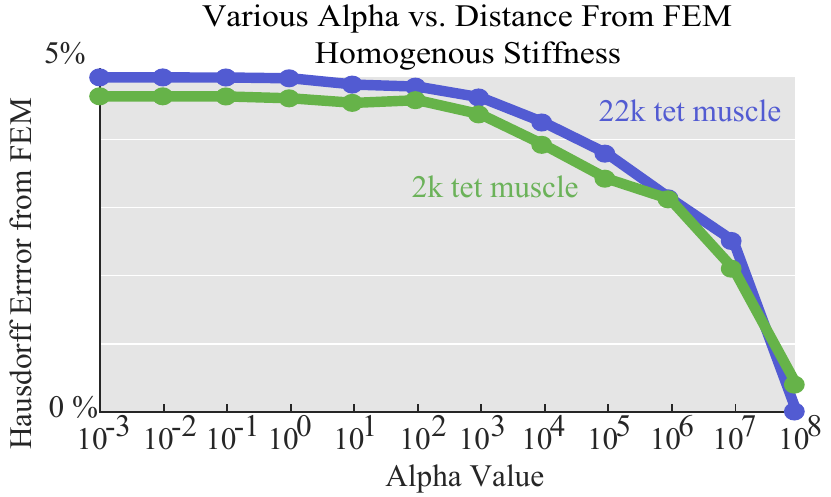}
    \caption{ Measures the Hausdorff distance divided by the rest length of the mesh between EMU results varying $\alpha$ values and the FEM result on a 2k tet muscle and 22k tet muscle with homogenous stiffness. As $\alpha$ increases the distance goes to 0.}
    \label{fig:alpha-plot}
\end{figure}

With all the pieces of EMU in place, we can now detail how we choose the ACAP energy weighting term, $\alpha$. On homogeneously stiff muscle meshes, higher $\alpha$ values produce deformations closer to FEM (\autoref{fig:alpha-plot}). However, since higher $\alpha$ increases the stiffness of the system, it non-linearly increases the number of newton iterations to convergence. For example, the 22k homogenous muscle mesh at $\alpha=1$ requires 21 newton iterations to converge while at $\alpha=1e8$ it requires 42,151 iterations. On the other hand, an exceedingly small $\alpha$ will allow the continuous mesh to drift away from the deformation gradients. 

With the addition of stiff tendons and bones into the system, the relationship between $\alpha$ and distance from FEM deformations is not as clear, as shown in \autoref{table:admm-FEM-EMU-table}. Under an exceedingly high $\alpha$ the stiff region's continuous mesh will not be allowed to deviate from the deformation gradient. As explained in \autoref{sec:affine-bones} and shown in \autoref{fig:muscle-ruler} this will lock the deformation since the bone vertices will not be allowed to deviate from the bone $F$ during the optimization. However, experiments show that there exists a value of $\alpha$ which results in deformations that resemble FEM even more closely than ADMM. Therefore, we provide a heuristic below to find a good $\alpha$.

\begin{table}[ht]
    \begin{center}
    \begin{tabular}{l r r}
    \hline
    \rowcolor{black!20}
    Tets & 12,184 & 51,271 \\
    \hline
     & \textit{Error} & \textit{Error} \\
    \hline
    \rowcolor{black!20}
    FEM & 0.000 & 0.000 \\
    \hline
    ADMM & 7.048 & 7.381\\
    \hline
    \rowcolor{black!20}
    EMU $\alpha=1e0$ & 12.095 & 11.119 \\
    \hline
    EMU $\alpha=1e1$ & 8.214 & 6.428\\
    \hline
    \rowcolor{black!20}
    EMU $\alpha=1e2$ & 2.143 & \cellcolor{black!40}1.500 \\
    \hline
    EMU $\alpha=1e3$ & 2.381 & 3.214\\
    \hline
    \rowcolor{black!20}
    EMU $\alpha=1e4$ & \cellcolor{black!40}1.667 & 3.119\\
    \hline
    EMU $\alpha=1e5$ & 8.095 & 5.786\\
    \hline
\end{tabular}
\end{center}
\caption{ Accuracy comparisons between ADMM, EMU and FEM with muscle activation under gravity. The error measured here is the Haussdorf distance from FEM for each result, divided by the rest-state length of the mesh.}
\label{table:admm-FEM-EMU-table}
\end{table}

\begin{figure}[H]
    \includegraphics[width=\columnwidth]{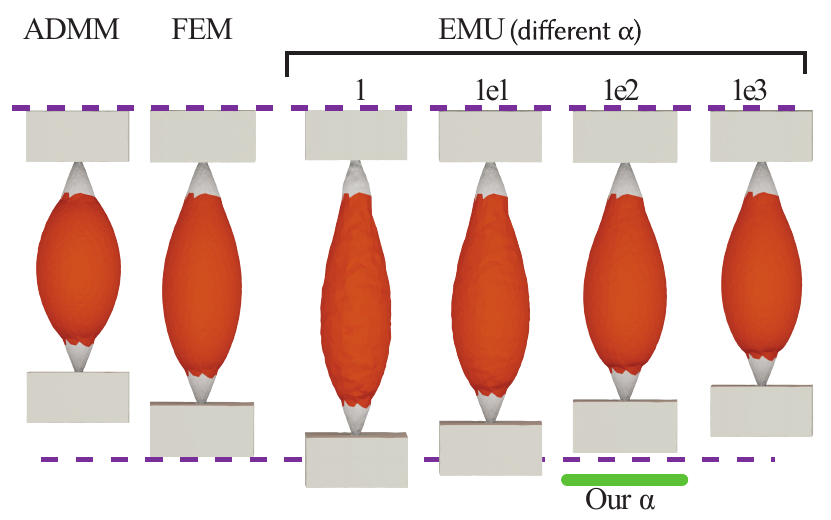}
    \caption{ Comparison of ADMM, FEM and EMU on fusiform muscles hanging under gravity. EMU results vary with the chosen alpha parameter but the optimal alpha chosen using our heuristic produces excellent agreement with the FEM solution and is more accurate than ADMM.} 
    \label{fig:compareADMM}
\end{figure}

To find a good $\alpha$ we linearly search over the 1D space of weighting parameters (starting with $\alpha=1$) and tally the number of newton iterations taken by EMU for some fixed muscle activation. We increase $\alpha$ until we observe a sharp increase in the number of Newton terations per step. We take the alpha immediately before this increase motivated by notion that penalty term optimizations admit an optimal alpha that exhibits fast convergene to the local optimum \cite{Wright1999}. We have found this heuristic generates good visual agreement with finite element results and also optimizes for speed as shown in \autoref{fig:compareADMM}. Let us note that our experiments illustrate that EMU produces visually pleasing results (though with differing deformations) for all values of $\alpha$. Ultimately, animation involves a fair bit of artwork, and visual appeal is subjective. Alpha can be used to tailor the visual output of the EMU simulation in more artistic applications.

\begin{table}[h!]
    \begin{center}
    \begin{tabular}{lrrr} 
    \hline 
    Model & Num Tets & Bones & Muscles \\ [0.5ex] 
    \hline
    \rowcolor{black!20} simple fusiform muscle & 600k & 2 & 1  \\ 
    simple bipennate muscle & 12k & 2 & 1 \\
    \rowcolor{black!20} simple contacting muscles & 20k & 2 & 2 \\
    curved contacting muscle & 11k & 2 & 1 \\
    \rowcolor{black!20} elephant head & 29k & 2 & 2 \\
    bicep & 33k & 3 & 1 \\
    \rowcolor{black!20} leg & 43k & 4 & 5 \\
    upper arm & 33k & 4 & 6 \\
    \rowcolor{black!20} soft robot wheel & 3k & 2 & 2 \\
    soft robot hand & 18k & 7 & 9 \\
    \rowcolor{black!20} cartoon skull & 48k & 2 & 4 \\
    chest-arm-back & 47k & 5 & 11 \\
    \rowcolor{black!20} Fruit picker & 289k & 6 & 6 \\
    Quadropus & 220k & 5 & 8 \\
    [1ex]
    \hline
    \end{tabular}
    \end{center}
    \caption{Musculoskeletal models simulated with EMU.}
    \label{table:stats}
\end{table}

\section{Results and Discussion}

We simulate a variety of musculoskeletal geometries using EMU. The tetrahedral count of our models range from a small 3k tetrahedron soft robot to a large 600k tetrahedron muscle as shown in \autoref{table:stats}. Not included in the table below were the various sized simple fusiform meshes generated for the performance and scalability tests. For each mesh, the initial step involved finding the first $k$ modes of the Hessian~(\refequ{Hessian-woodbury}) once, upfront. The runtime of this pre-processing step differs significantly based on the mesh connectivity and number of bones in the mesh, but ranges from several seconds to several minutes for our larger examples. The second pre-processing step involves choosing an $\alpha$. This involves, at most, 10 Newton solves on the mesh. However, since these are one time operations, we exclude them from our performance numbers.

\begin{figure*}[htp]
\includegraphics[width=\textwidth]{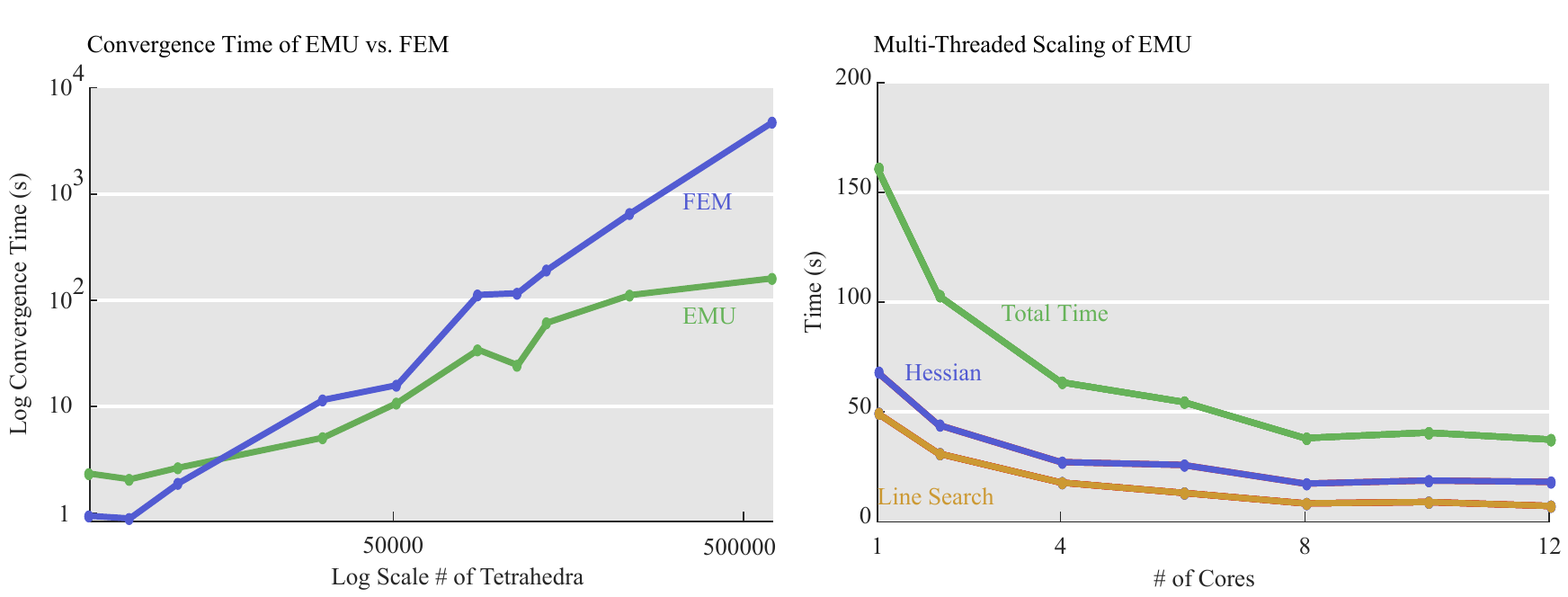}
\caption{Left: single core scaling of EMU and standard FEM as a function of mesh size. Right: multi-core scaling. EMU is faster than standard FEM for all but the smallest meshes and exhibits better asymptotic behavior. The EMU quasi-static solve scales well with number of cores. When using 12 cores the bottleneck in the code is the sparse back solve required for ACAP energy calculation which could be optimized further using more efficient solvers.}  
\label{fig:perF}
\end{figure*}

EMU exhibits excellent performance when compared to the state-of-the-art open source FEM solver ~\cite{gauss2019}. Our FEM algorithm uses Stable Neo-Hookean elasticity from \cite{Smith2018} solved using the open source Pardiso solver, \cite{pardiso-6.0a,pardiso-6.0b,pardiso-6.0c}. We compare the performance of both algorithms in terms of scaling with respect to mesh size and scaling with respect to number of available CPU cores. For testing we measured convergence of both methods by checking if the change in energy of an iteration was $<1e-2$. We found this sufficient to produce results with excellent visual fidelity. Single-core scaling tests were performed on Intel Core i7-6700HQ CPU (2.60GHz). Multi-core tests were performed on a Dual Intel Xeon Gold 5120 (2.20GHz). Scaling tests were done on our simple fusiform muscle (\autoref{fig:perF}). Every mesh, including the 600k tetrahedral mesh ran without memory issues on a 16GB RAM laptop. Since we only use fixed size dense matrices, memory usage is not a problem in our simulations.

On a single core machine, EMU scales better than state-of-the-art FEM, as a function of number of mesh tetrahedra (\autoref{fig:perF}). Our most intensive operations are a sparse back substitution required to solve \refequ{acap} and a dense matrix inversion required to compute the Woodbury identity. For a given example, increasing mesh resolution does not have a large effect on the spectral characteristics of the Hessian in~\refequ{acap}. Thus $k$~(\refequ{Hessian-woodbury}) typically stays constant so the cost of the required dense solve remains fixed. The effect of this is that EMU is faster than FEM for all but the smallest examples and, for medium to large meshes, impressively so -- exhibiting speedups of over $20\times$. This is particularly impressive when one considers that EMU is not reducing the solution space in any way; it is solving the same problem as the FEM discretization. Even though not specifically designed for simulating isotropic, homogenous materials, EMU retains its performance advantage over FEM. We tested the performance on the Stanford Bunny mesh, up to 40k tetrahedra and observed a \emph{5-6x} speed-up over state-of-the-art FEM solver. 

EMU also parallelizes well. We observe a further \emph{3x} performance improvement by running EMU on a 12 core machine (hyperthreading disabled). As discussed, the EMU Hessian update is extremely parallizable. The bottleneck in our implementation comes from the limited ability of the linear solver (Pardiso) to efficiently parallelize the backsolves needed to minimize the ACAP energy. Exploiting other solvers with better parallel scaling would improve EMU performance even further. 

\begin{figure}[H]
    \includegraphics[width=\columnwidth]{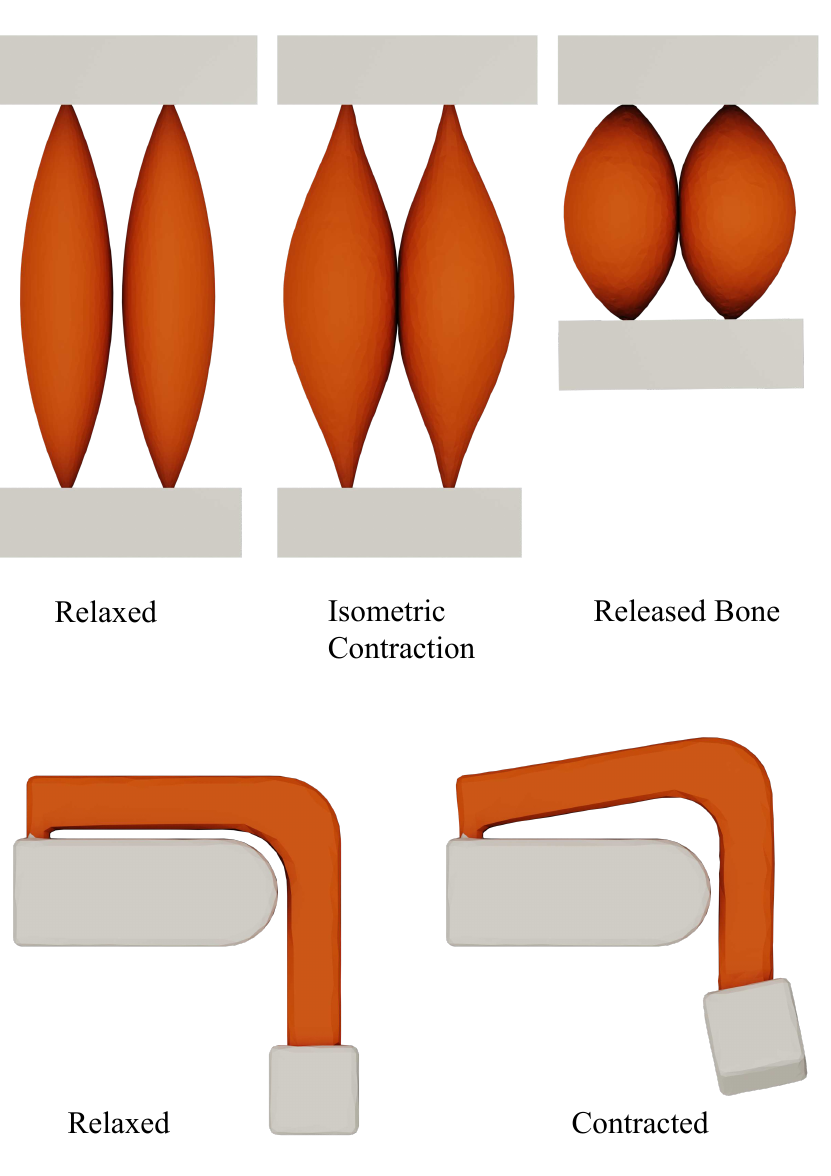}
    \caption{ Examples of muscles in contact, simulated using EMU. Top: EMU prevents side-by-side muscles from interpenetrating under both isometric, and full motion contraction. Bottom: A muscle with a sharp bend pulls a square bone around a corner. (Video: 1m05s, 1m50s)} 
    \label{fig:contact}
\end{figure}

Our contact-aware solver is able to simulate muscles in close contact and allows EMU to exert force along relatively complex muscle paths. The top row of \autoref{fig:contact} shows two fusiform muscles in a side-by-side configuration. The muscles are isometrically contracted (the bottom bone is fixed) and then the bone is released allowing the muscles to fully contract. Our contact solver prevents interpenetrations in both cases. The bottom row of \autoref{fig:contact} shows a muscle with a sharp bend. Contracting the fibers in this muscle allows it to exert force around a corner and apply a vertical force to the square bone at its end-point. Again, the EMU contact solver prevents the muscle from contracting into the underlying bones. 

\begin{figure}[h]
    \includegraphics[width=\columnwidth]{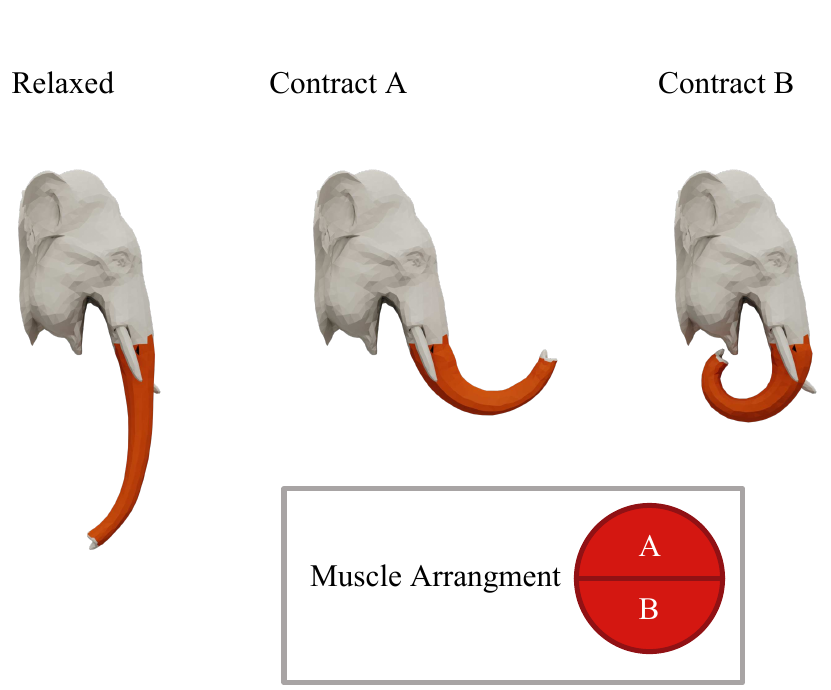}
    \caption{ A simulated elephant trunk. EMU can efficiently generate realistic, large deformation motions of this elephant trunk by contracting the appropriate muscles. Bottom: Muscle arrangement of the trunk. Muscle fibers run along the trunks length. (Video: 2m05s)} 
    \label{fig:elephant}
\end{figure}

\begin{figure}[h]
    \includegraphics[width=\columnwidth]{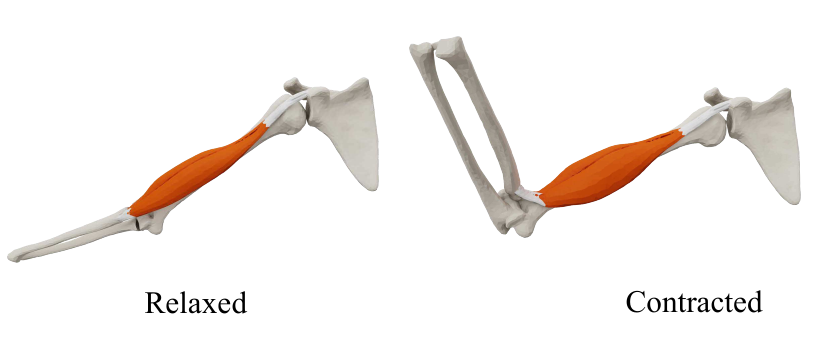}
    \caption{ Simulating the motion of the humerus, radius and ulna induced by contracting the bicep. The bicep is a biarticular muscle which connects the shoulder directly to the forearm, skipping the humerus entirely. EMU can handle such complicated muscles } 
    \label{fig:bicep}
\end{figure}

Next we show that EMU can generate realistic large scale muscle-first motions by simulating an elephant trunk~(\autoref{fig:elephant}). The cross-section of the trunk is divided into quarters with each quarter being an independent muscle. Fibers run along the length of the trunk and contracting various muscles causes the trunk to bend. We simulate the canonical elephant feeding motion -- the elephant reaches up to grab food, then bends the trunk in the opposite direction to bring the food to its mouth.

 \begin{figure}[h]
    \includegraphics[width=\columnwidth]{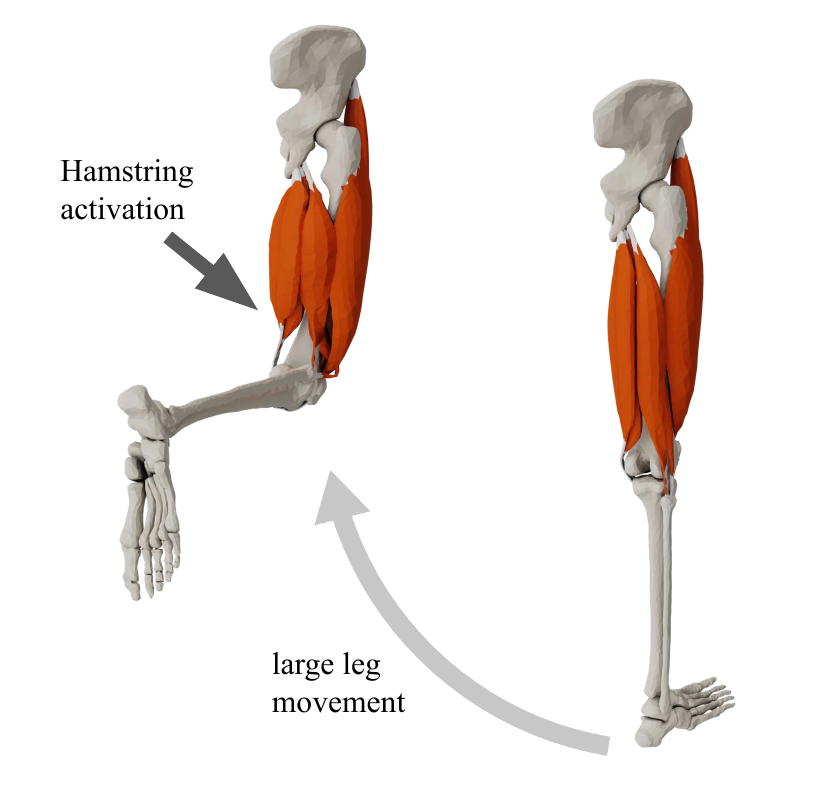}
    \caption{Simulating the motion of a leg, induced by contracting the hamstring. (Video: 0m41s)} 
    \label{fig:leg}
\end{figure}

One of the advantages of the EMU deformation gradient formalism is the ease with which joints can be incorporated. \autoref{fig:bicep} shows how the contraction of the bicep drives the motion of the humerus, radius and ulna. This is because the bicep is biarticular -- it connects the shoulder directly to the forearm. Motion of the humerus results from the forearm being driven by the muscle. EMU effortlessly handles complicated muscles such as this and is able to properly transmit forces from the contracting muscle and through both the elbow (modelled as a hinge joint) and the shoulder (modelled as a spherical joint). \autoref{fig:leg} shows a simulation of a contracted hamstring which drives the large scale motion of a multi-muscle leg. This shows the ability of EMU to generate realsitic, muscle-first motion in the presence of multiple muscles, joints, tendons and bones. 

\begin{figure}[h]
    \includegraphics[width=\columnwidth]{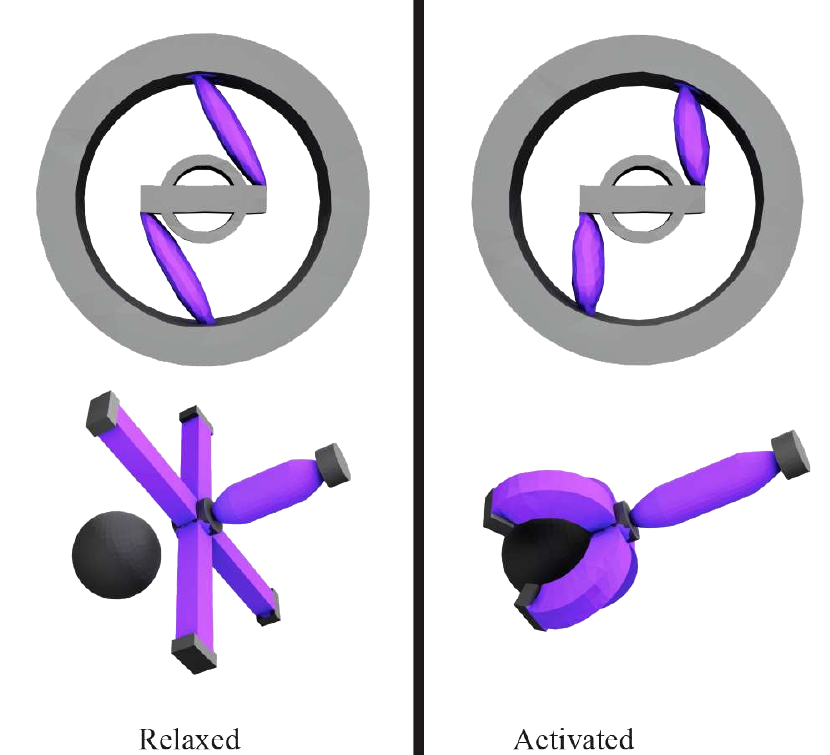}
    \caption{ EMU can also be used to simulate pneumatically actuated mechanical systems. Here we use EMU's ability to simulate deformable objects, joints and rigid bodies, to simulate (Top) this soft mechanism that can rotate its outer ring when its actuators are contracted and (Bottom) a soft robotic grasper. (Video: 2m25s)}
    \label{fig:robot}
\end{figure}

\begin{figure}[h]
  \includegraphics[width=\columnwidth]{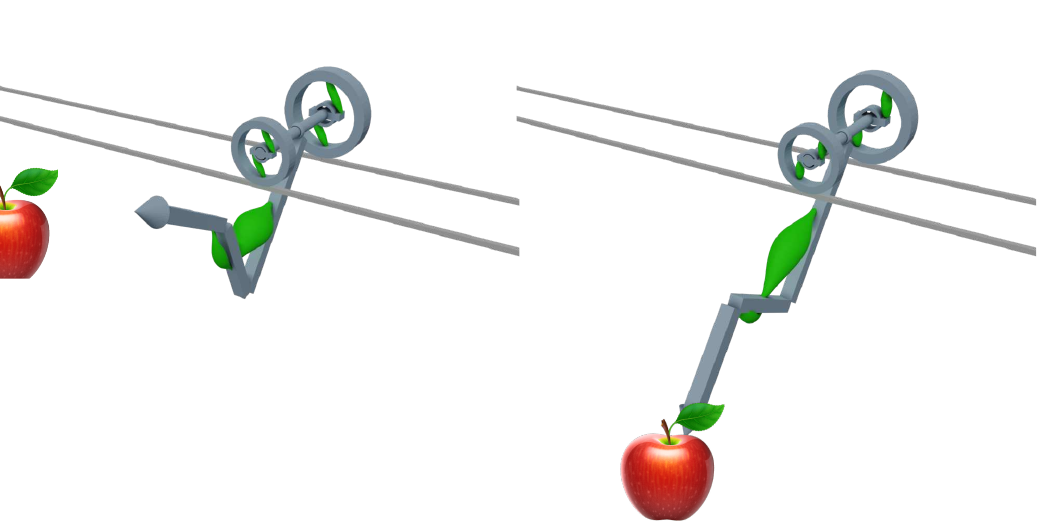}
    \caption{A 289k tetrahedron fruit picking robot which rolls back and forth along the rails, grabs fruits and drops them.(Video: 2m19s)}
    \label{fig:bigsoftrobot}
\end {figure}

\begin{figure}[h]
    \includegraphics[width=\columnwidth]{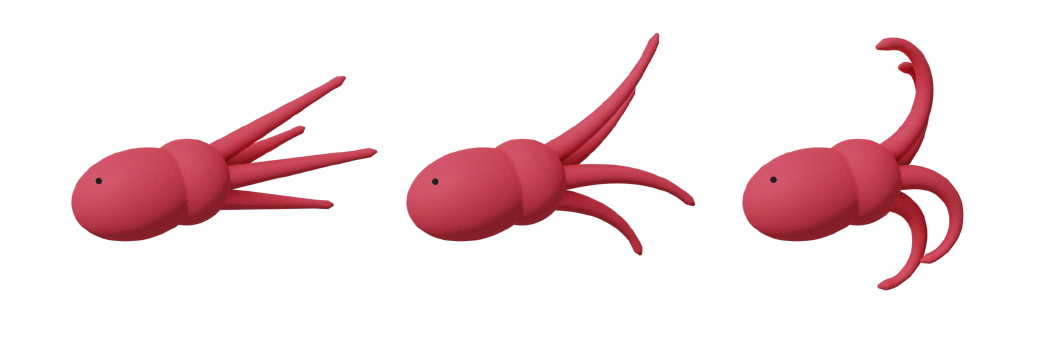}
    \caption{EMU used to simulate the motion of a squid-like animal with 220k tetrahedrons. (Video: 0m0s)}
    \label{fig:Quadropus}
\end{figure}

EMU has uses beyond biomenchanical simulation. In \autoref{fig:robot} (Top) we simulate a pneumatically actuated mechanism. This mechanism can rotate its outer ring when its pneumatic actuators are activated. EMU's ability to handle deformable and rigid bodies, connected with joints, is perfect for such applications. \autoref{fig:robot} (Bottom) shows the simulation of a soft robotic gripper which reaches down and grasps a ball. Next, we present two more examples. In \autoref{fig:bigsoftrobot}, a robot which can pick ripe fruits for juicing simulated by 289k tets at 13.3 seconds per frame. And in \autoref{fig:Quadropus} we present a squid-like creature discretized by 200k tets simulated at 127.9 seconds per frame.

\begin{figure}[h]
    \includegraphics[width=\columnwidth]{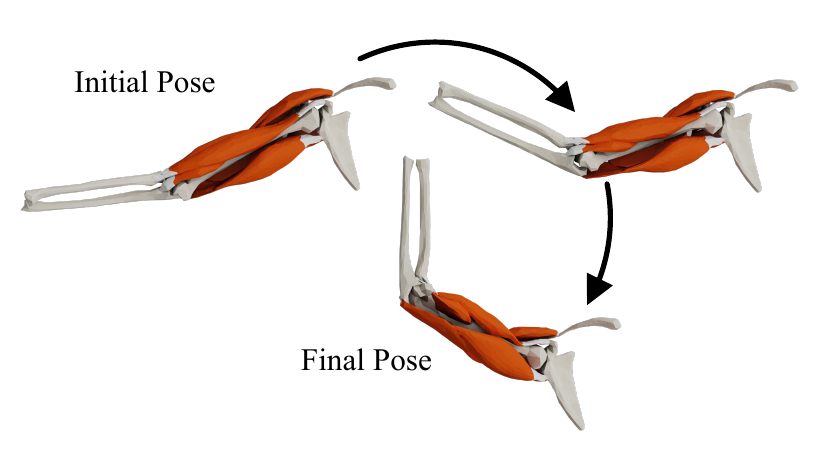}
    \caption{ The skeletal motion of this arm is scripted by an artist and we use EMU to  add compelling muscle deformations. (Video: 2m12s)} 
    \label{fig:scripted}
\end{figure}

\begin{figure}[h]
    \includegraphics[width=\columnwidth]{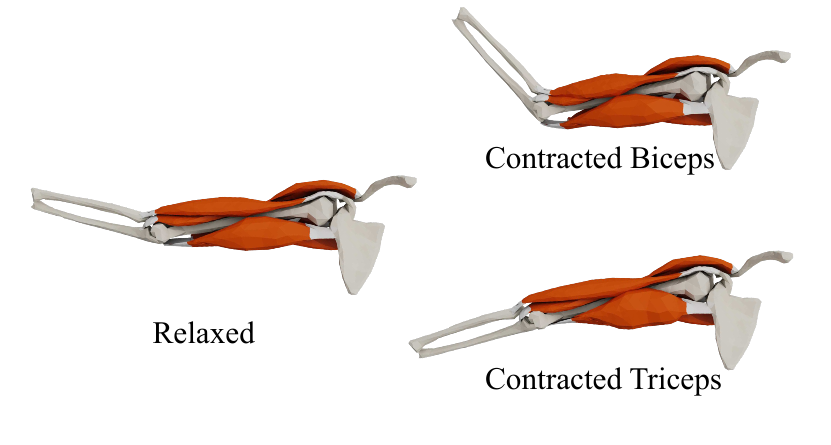}
    \caption{ A simulation using a full musculoskeletal model of a human arm. We use EMU to simulate a bicep contraction, followed by a tricep contraction to hyper-extend the elbow. (Video: 1m56s)} 
    \label{fig:bicep_tricep}
\end{figure}

\begin{figure*}[ht]
    \includegraphics[width=\textwidth]{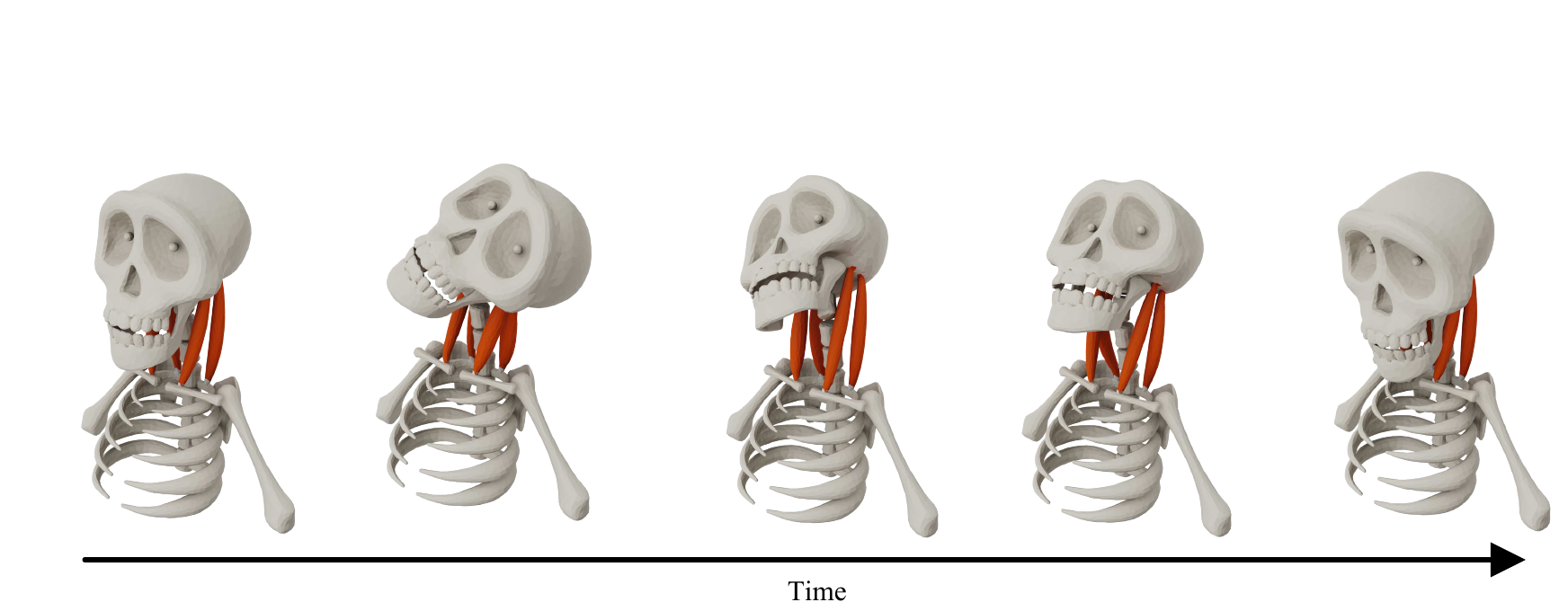}
    \caption{ This cartoon head roll is completely driven by actuating the four neck muscles of this model. The neck joint is hollistically simulated using EMU. (Video: 0m08s)} 
    \label{fig:head_roll}
\end{figure*}

Finally we turn our attention to more complex biomechanical models. \autoref{fig:bicep_tricep} demonstrates EMU's ability to generate simulations using realisitc biomechanical activation sequences. In this figure we simulate a bicep contraction, followed by a tricep contraction. This first flexes the elbow and then hyperextends. This example shows off all of EMUs features, its efficiency, and its ability to seamlessly animate bones, tendons and muscles to generate muscle-first bulk musculoskeletal motion. \autoref{fig:head_roll} illustrates the use of EMU to generate muscle-first head motion. This cartoon head roll is completely driven by muscle actuations of the four neck muscles. EMU allows us to flex the muscles of this complex upper body model (\autoref{fig:teaser}) to get a biomechanically feasible pump. Finally we can also use EMU to add muscle motion on top of scripted bones, like this arm mesh~(\autoref{fig:scripted}) which is animated to rotate through a large motion, and creates natural looking complimentry muscle motion. 

\section{Conclusion and Future Work}

We have presented a new, efficient algorithm for bulk musculoskeletal simulation. Our algorithm is a multi-objective, discontinuous iterative approach to finite element simulation which uses a novel, minimal energy penalty to enforce continuity. We demonstrate how this approach leads to the construction of an efficient algorithm for musculoskeletal simulation which at run time requires only the inversion of a small dense matrix, sparse back substitution and the factorization of a block diagonal matrix (all extremely fast operations). 

Furthermore we show how to incorporate both bones and simple tendons into the method without needing to resort to specialized approaches such as coupled rigid body simulations or line-of-force methods. To our knowledge, we are the first to demonstrate such a holistic approach to bulk musculoskeletal simulation (as all previous approaches avoid tendonous attachment points as in \autoref{fig:bicep} and \autoref{fig:leg}).

Although EMU has performance benefits over FEM and our results show visually appealing deformations, for simulations where accuracy is essential, EMU falls short of FEM due to the inexactness introduced by the $\alpha$ parameter. However, we show that by using our heuristic to find an optimal $\alpha$, we can better approximate the deformations produced by FEM than other contemporary methods such as ADMM. Additionally, unlike ADMM and other projection based methods, EMU allows the use of any material model and muscle activation model.

We believe our method will have immediate application for character animations and control, but we are most excited about the areas of future work EMU opens up, both on the numerical optimization front and in biomechanical simulations. Incorporating more complicated tendon routing and sliding is a crucial area of future work. Previous approaches for efficient sliding work on 1D~\cite{Sueda2011} and 2D~\cite{Weidner2018} geometries, but volumetric tendon sliding is unaddressed. Co-dimensional simulation could also be explored, allowing the coupling of 1D, 2D and 3D elements inside of our framework. Since we only need to represent the deformation of such elements our method should be well suited to this. We could also improve the performance of EMU by exploring the use of fast solvers in our continuity energy (akin to using specialized solvers in the pressure projection step of a fluid simulation). 

Modeling is a crucial area of future work that would benefit from the availability of an algorithm like EMU. Idealized joint constraints restrict the motions that can be achieved by any simulator, no matter how fast or how robust. Using high-performance simulators like EMU will enable us to model connections between bones using ligaments and other soft tissue structures, thus capturing more natural motions. Building tools that can produce these detailed volumetric representations of the human body will be increasingly important moving forward. 

\section*{Acknowledgements}
This work is funded in party by National Science Foundation (CAREER-1846368), NSERC Discovery (RGPIN-2017-05524, RGPIN-2017-05235, RGPAS-2017-507938), Connaught Fund (503114), CFI-JELF Fund, Accelerator (RGPAS-2017-507909), New Frontiers of Research Fund (NFRFE–201), the Ontario Early Research Award program, the Canada Research Chairs Program, the Fields Centre for Quantitative Analysis and Modelling and gifts by Adobe Systems, Autodesk and MESH Inc. We thank John Kanji, and Josh Holinaty for help with designing figures; Rinat Abdrashitov for his video editing skills; Josh Holinaty for lending us his beautiful voice in the video; Sarah Kushner, Honglin Chen, Abhishek Madan, Hsueh-Ti Derek Liu and Darren Moore for proofreading; John Hancock for IT support; anonymous reviewers for their helpful comments and suggestions.
\bibliographystyle{eg-alpha-doi} 
\bibliography{references}       


\end{document}